# Testing a Bell Inequality with a Remote Quantum Processor

Jed Brody, Emory University, Atlanta, GA

Robert Avram, Saint Ann's School, Brooklyn, NY

IBM Quantum offers free remote access to real quantum processors.[1] One of the many experiments now accessible to all students is a test of Bell inequalities. This experiment introduces the rigorous mysteries that physicists have grappled with for a century. Using IBM Quantum to test Bell inequalities is not new.[2,3] However, we are unaware of any single reference, appropriate for introductory students, that contains (1) the derivation of a Bell inequality, (2) the derivation of the corresponding quantum prediction, and (3) instructions for carrying out the experiment with IBM Quantum.

A Bell inequality is a mathematical constraint imposed by common sense. Our common-sense assumption is called local realism: The measurement of one object can't affect an arbitrarily distant object (locality), and the properties of objects exist regardless of whether they've been measured (realism).

## A Quick Derivation of a Bell Inequality

One of the easiest Bell inequalities to derive is the Clauser-Horne-Shimony-Holt (CHSH) inequality.[4-7] We imagine a system consisting of two objects, A and B, that are measured separately, as shown in Figure 1. Each measurement device has two settings, labeled 1 and 2. We do not need to specify the nature of the objects or the measurements, except that each measurement yields one of two results, which we will designate as +1 and -1. These two numbers are arbitrary and dimensionless, analogous to assigning +1 to "heads" and -1 to "tails" when recording the results of a coin toss. Similarly, if our measurement device is a Stern-Gerlach apparatus,[8] +1 could indicate "deflected toward the north pole of the magnet (spin up)," -1 could indicate "deflected toward the south pole of the magnet (spin down)," and the measurement settings control the orientation of the magnet.

The measured result ($\pm 1$) for object A is called $A_1$ when the corresponding measurement setting is 1, and the measured result is called $A_2$ when the corresponding measurement setting is 2. $B_1$ and $B_2$ are defined similarly for the measurements of object B.

For each pair of objects A and B, we choose one measurement setting for object A (so we're measuring either $A_1$ or $A_2$ but not both), and we choose one measurement setting for object B (so we're measuring either $B_1$ or $B_2$ but not both). Suppose we measure $A_1$ and $B_1$ for many pairs of objects. For each pair of objects, we choose to calculate the product $A_1B_1$, which must be $\pm 1$. We can then find the average of $A_1B_1$. Similarly, we can find the averages of of $A_1B_2$, $A_2B_1$, and $A_2B_2$.

Next, we define the quantity

$S = A_1B_1 - A_1B_2 + A_2B_1 + A_2B_2.$ (1)

S does not have any obvious physical significance; it's just something we can calculate if we know all four variables on the right side of the equation. Since each of these four variables is ±1, S must be ±2: S = $A_1(B_1 - B_2) + A_2(B_1 + B_2)$, and one of the quantities in parentheses must 0 and the other must be ±2, and the variables in front of the parentheses are ±1.

Since S must be ±2, the average of S for many pairs of objects must be between -2 and +2: $|S_{average}| \leq 2$. This is the CHSH inequality. Restated in terms of Eq. (1), and using angle brackets to represent averages,

$$|<S>| = |<A_1B_1> - <A_1B_2> + <A_2B_1> + <A_2B_2>| \qquad (2)$$

must be less than or equal to 2. The assumption of local realism is subtle: S depends on both $A_1$ and $A_2$, so Eq. (1) implicitly assumes that $A_1$ and $A_2$ both exist for every object A, even though only one of the two variables is measured. We also implicitly assume that $A_1$ and $A_2$ do not depend on the measurement setting for object B.

## The Quantum Prediction

In contrast with local realism, quantum theory predicts that the CHSH inequality can be disobeyed. To disobey the CHSH inequality, we need quantum entanglement. Quantum entanglement is easy to create in IBM Quantum.

A classical binary digit (bit) is either 0 or 1. A variable that can be either 0 or 1 is called a Boolean variable. A quantum bit (qubit) can be in a superposition, written $a|0\rangle + b|1\rangle$.[9] When the qubit is measured, the result obtained is either 0, with probability $|a|^2$, or 1, with probability $|b|^2$. 0 and 1 are Boolean labels for the two possible physical states of the qubit after measurement. We can think of $|0\rangle$ and $|1\rangle$ as unit vectors; they are not numbers, but they are multiplied by scalars (a and b) indicating their importance to the total state.

We want to entangle two qubits. When we have two qubits, there are four possible measurement results: 00, 01, 10, and 11. Prior to measurement, the state may have the form $a|0\rangle|0\rangle + b|0\rangle|1\rangle + c|1\rangle|0\rangle + d|1\rangle|1\rangle$, where the squared norms of the coefficients are the probabilities of measuring the corresponding results. We can think of $|0\rangle|0\rangle$, $|0\rangle|1\rangle$, $|1\rangle|0\rangle$, and $|1\rangle|1\rangle$ as the four unit vectors of a two-qubit system.

By default, the starting state in IBM Quantum is $|0\rangle|0\rangle$: If the qubits are measured at this point, the result is certain to be 00. We manipulate the state by applying "quantum gates." We need only two gates to entangle the qubits.

We first apply something called a Hadamard gate, H, to the first qubit. The effect of H on $|0\rangle$, shown in Fig. 2, is to create a superposition: $H|0\rangle = \frac{1}{\sqrt{2}}(|0\rangle + |1\rangle)$. The state of the two qubits is now $\frac{1}{\sqrt{2}}(|0\rangle + |1\rangle)|0\rangle = \frac{1}{\sqrt{2}}(|0\rangle|0\rangle + |1\rangle|0\rangle)$, writing the top qubit first. Next we apply the controlled NOT gate, CNOT, represented by the rightmost symbol in Fig. 3. CNOT has no effect if the top qubit is $|0\rangle$. If the top qubit is $|1\rangle$, the bottom qubit flips (from $|0\rangle$ to $|1\rangle$ or from $|1\rangle$ to $|0\rangle$). The CNOT acts separately on the two terms in our state, so $|0\rangle|0\rangle$ is unaffected, but $|1\rangle|0\rangle$ changes to $|1\rangle|1\rangle$. Now our state is $\frac{1}{\sqrt{2}}(|0\rangle|0\rangle + |1\rangle|1\rangle)$. This is an entangled state: It cannot be factored into a product of the states of the two individual qubits.

Next, we need to establish two ways to measure each of two qubits so that we can perform measurements of $A_1$ and $A_2$ (on the top qubit), and of $B_1$ and $B_2$ (on the bottom qubit). Regardless of how a qubit is made, $|0\rangle$ can model "spin up," and $|1\rangle$ can model "spin down." A detailed familiarity with spin is unnecessary. We need to know only that spin is a property that can be measured in different directions. By convention, $|0\rangle$ represents spin in the +z direction, and $|1\rangle$ represents spin in the –z direction, as shown in Fig. 4. Figure 4 shows the xz-plane of the Bloch sphere,[9] and the vectors are known as Bloch vectors. To represent spin in an arbitrary direction in the xz-plane, we can use the qubit $|\theta\rangle = \cos(\theta/2)|0\rangle + \sin(\theta/2)|1\rangle$. Students can confirm that $|\theta=0°\rangle = |0\rangle$ and $|\theta=180°\rangle = |1\rangle$.

Every measurement in IBM Quantum yields 0 or 1, representing a measurement of spin along the z axis. To effectively measure spin in any other direction in the xz-plane, we need to rotate the Bloch vector from the direction we want, onto the z axis. The $R_y(-\theta)$ gate accomplishes this, where the subscript indicates rotation around the y axis perpendicular to Fig. 4. $R_y(-\theta)$ affects $|0\rangle$ and $|1\rangle$ as follows:

$$R_y(-\theta)|0\rangle = \cos(\theta/2)|0\rangle - \sin(\theta/2)|1\rangle, \qquad (3)$$

and

$$R_y(-\theta)|1\rangle = \sin(\theta/2)|0\rangle + \cos(\theta/2)|1\rangle. \qquad (4)$$

Positive θ represents clockwise rotation in Fig. 4, which is why we apply $R_y(-\theta)$ to rotate counterclockwise from the measurement direction onto the z axis. Students can confirm $R_y(-\theta)|\theta\rangle=|0\rangle$, where $|\theta\rangle = \cos(\theta/2)|0\rangle + \sin(\theta/2)|1\rangle$.

To measure the first qubit along direction $\alpha$ and the second qubit along direction $\beta$, we apply $R_y(-\alpha)$ to the first qubit and $R_y(-\beta)$ to the second qubit. This means that our entangled state, $\frac{1}{\sqrt{2}}(|0\rangle|0\rangle + |1\rangle|1\rangle)$, becomes $\frac{1}{\sqrt{2}}[R_y(-\alpha)|0\rangle R_y(-\beta)|0\rangle + R_y(-\alpha)|1\rangle R_y(-\beta)|1\rangle]$. Using Eqs. (3) and (4), we obtain $\frac{1}{\sqrt{2}}[(\cos\frac{\alpha}{2}|0\rangle - \sin\frac{\alpha}{2}|1\rangle)(\cos\frac{\beta}{2}|0\rangle - \sin\frac{\beta}{2}|1\rangle) + (\sin\frac{\alpha}{2}|0\rangle + \cos\frac{\alpha}{2}|1\rangle)(\sin\frac{\beta}{2}|0\rangle + \cos\frac{\beta}{2}|1\rangle)]$. Multiplying binomials, while taking care to distinguish $|0\rangle|1\rangle$ from $|1\rangle|0\rangle$, we obtain $\frac{1}{\sqrt{2}}[(\cos\frac{\alpha}{2}\cos\frac{\beta}{2}|0\rangle|0\rangle - \cos\frac{\alpha}{2}\sin\frac{\beta}{2}|0\rangle|1\rangle - \sin\frac{\alpha}{2}\cos\frac{\beta}{2}|1\rangle|0\rangle + \sin\frac{\alpha}{2}\sin\frac{\beta}{2}|1\rangle|1\rangle) + (\sin\frac{\alpha}{2}\sin\frac{\beta}{2}|0\rangle|0\rangle + \sin\frac{\alpha}{2}\cos\frac{\beta}{2}|0\rangle|1\rangle + \cos\frac{\alpha}{2}\sin\frac{\beta}{2}|1\rangle|0\rangle + \cos\frac{\alpha}{2}\cos\frac{\beta}{2}|1\rangle|1\rangle)]$. Combining terms with the same "unit vector" and simplifying with trigonometric identities yields $\frac{1}{\sqrt{2}}(\cos\frac{\alpha-\beta}{2}|0\rangle|0\rangle + \sin\frac{\alpha-\beta}{2}|0\rangle|1\rangle + \sin\frac{\beta-\alpha}{2}|1\rangle|0\rangle + \cos\frac{\alpha-\beta}{2}|1\rangle|1\rangle)$.

The probability of each result is found by squaring the corresponding coefficient in the final expression above. So the probabilities of measuring the results 00, 01, 10, and 11 are

$$P(00) = \frac{1}{2}\cos^2\frac{\alpha-\beta}{2}, \qquad (5)$$

$$P(01) = \frac{1}{2}\sin^2\frac{\alpha-\beta}{2}, \qquad (6)$$

$$P(10) = \tfrac{1}{2}\sin^2\tfrac{\alpha-\beta}{2}, \tag{7}$$

and

$$P(11) = \tfrac{1}{2}\cos^2\tfrac{\alpha-\beta}{2}, \tag{8}$$

respectively.

In the derivation of the CHSH inequality, the two possible measurement results are +1 and -1, not 0 and 1. To match the assumptions made in the derivation of the CHSH inequality, we must use Table 1 to map the Boolean labels (0 and 1) to the "spin" values used in calculations (±1). The product of the two spin values is +1 for 00 and 11, and the product of the two spin values is -1 for 01 and 10. So P(00)+P(11) is the probability that the product is +1, and P(01)+P(10) is the probability that the product is -1. Finally, <AB>, the average of the product of the two values, is

$$<AB> = (+1) \times P(00) + (-1) \times P(01) + (-1) \times P(10) + (+1) \times P(11). \tag{9}$$

Combining Eqs. (5)-(9),

$$<AB> = \tfrac{1}{2}\cos^2\tfrac{\alpha-\beta}{2} - \tfrac{1}{2}\sin^2\tfrac{\alpha-\beta}{2} - \tfrac{1}{2}\sin^2\tfrac{\alpha-\beta}{2} + \tfrac{1}{2}\cos^2\tfrac{\alpha-\beta}{2}$$
$$= \cos^2\tfrac{\alpha-\beta}{2} - \sin^2\tfrac{\alpha-\beta}{2} = \cos(\alpha-\beta). \tag{10}$$

At long last, we can calculate <S> in the CHSH inequality. If $\alpha_1$, $\alpha_2$, $\beta_1$, and $\beta_2$ are the measurement angles for measurements of $A_1$, $A_2$, $B_1$, and $B_2$, respectively, then Eqs. (2) and (10) give the quantum prediction for <S>: $\cos(\alpha_1-\beta_1) - \cos(\alpha_1-\beta_2) + \cos(\alpha_2-\beta_1) + \cos(\alpha_2-\beta_2)$. If we choose $\alpha_1 = 0$, $\alpha_2 = \pi/2$, $\beta_1 = \pi/4$, and $\beta_2 = 3\pi/4$, then we find $<S> = 2\sqrt{2}$, contradicting the CHSH inequality, $|<S>| \leq 2$. So we have two conflicting predictions for <S>, and we need to do an experiment to see which, if either, is correct.

## Experiment: Which Theory Is Correct?

Using the IBM Quantum Composer, we constructed the circuit in Fig. 5 to determine $<A_1B_1>$. The symbols on the right represent measurements (along the z direction of the Bloch sphere). The Quantum Composer allows students to form quantum circuits simply by selecting gates from a graphical menu and arranging them as desired. The angle of rotation of the $R_y(-\theta)$ gate can be edited with a right click.

We selected the quantum processor called ibmq_jakarta and ran the circuit 1024 times, simply by specifying 1024 "shots." Each time the circuit ran, the measurement yielded either 00, 01, 10, or 11. IBM Quantum consolidated the results from the 1024 shots and listed the numbers of occurrences of 00, 01, 10, and 11: 413, 100, 90, and 421, respectively. We divided these four numbers by 1024 to estimate the probabilities P(00), P(01), P(10), and P(11). Next, we calculated $<A_1B_1>$ with Eq. (9). To get a standard deviation in $<A_1B_1>$, we repeated this sequence to determine $<A_1B_1>$ nine additional times.

This entire process was repeated using the circuits in Figs. 6-8 to determine means and standard deviations of $\langle A_1B_2\rangle$, $\langle A_2B_1\rangle$, and $\langle A_2B_2\rangle$.

We found $\langle A_1B_1\rangle = 0.655 \pm 0.017$, $\langle A_1B_2\rangle = -0.63 \pm 0.04$, $\langle A_2B_1\rangle = 0.638 \pm 0.014$, and $\langle A_2B_2\rangle = 0.62 \pm 0.03$. Equation (2) then gives $\langle S\rangle = 2.55 \pm 0.05$,[10] exceeding the CHSH inequality by 11 standard deviations. Our $\langle S\rangle$ is not as high as the quantum prediction of $2\sqrt{2}$ because systematic error tends to reduce $\langle S\rangle$.[11]

Educators are able to reserve a quantum processor so that students get their results within minutes. The experiment is quick, but the analysis requires time and concentration, and rumination over the philosophical implications can last a lifetime.

Table 1. Correspondence between Boolean label and dimensionless spin value. If the qubits actually consisted of spin-1/2 particles, we could multiply the dimensionless value by Planck's constant divided by $4\pi$ to obtain the correct units and magnitude of the z component of spin. We use the spin model for ease of visualization (on the Bloch sphere), but in fact, the IBM qubits consist of superconducting circuits.

| Spin Model | Boolean Label | Dimensionless Spin Value |
| --- | --- | --- |
| Spin Up | 0 | +1 |
| Spin Down | 1 | -1 |

**References**

1. https://quantum-computing.ibm.com/, with extensive tutorials at https://quantum-computing.ibm.com/composer/docs/iqx/.

2. D. Alsina and J.I. Latorre, "Experimental Test of Mermin Inequalities on a Five Qubit Quantum Computer," *Physical Review A* **94**, 012314 (July 2016).

3. D. García-Martín and G. Sierra, "Five Experimental Tests on the 5-Qubit IBM Quantum Computer," *Journal of Applied Mathematics and Physics* **6**, 1460 (July 2018).

4. A. Peres, "Unperformed experiments have no results," *American Journal of Physics* **46**, 745 (July 1978).

5. V. Scarani, *Quantum Physics: A First Encounter* (Oxford University Press, 2006).

6. D. Kaiser, *How the Hippies Saved Physics* (W. W. Norton & Company, 2011).

7. J. Brody, *Quantum Entanglement* (MIT Press, 2020).

8. The simulation at https://www.st-andrews.ac.uk/physics/quvis/simulations_html5/sims/quantum-versus-hv1/quantum-versus-hv1.html gives the user a choice of three orientations of each magnet.

9. W. Dur and S. Heusler, "Visualization of the Invisible: The Qubit as Key to Quantum Physics," *Phys. Teach.* **52**, 489 (Nov. 2014).

10. We propagated error using Eq. (3) in C. G. Deacon, "Error Analysis in the Introductory Physics Laboratory," *Phys. Teach.* **30**, 368 (Sept. 1992), but introductory students could use Eq. (2) therein.

11. D. Dehlinger and M. W. Mitchell, "Entangled photons, nonlocality, and Bell inequalities in the undergraduate laboratory," Am. J. Phys. **70**, 903 (Sept. 2002).

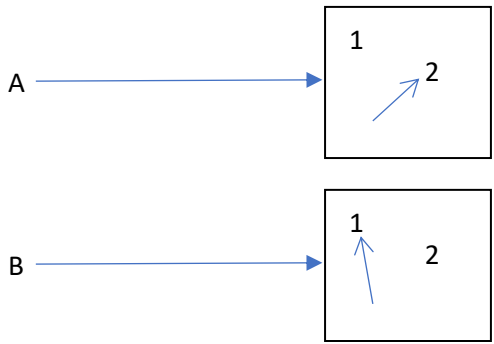

Figure 1.  A system of two objects, A and B, traveling into measuring devices, each with two switches. The switches are shown in position to measure $A_2$ and $B_1$.

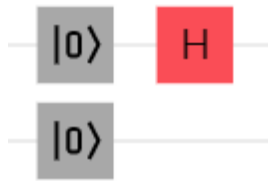

Figure 2.  The Hadamard gate (H) changes the top qubit from $|0\rangle$ to $\frac{1}{\sqrt{2}}(|0\rangle + |1\rangle)$.

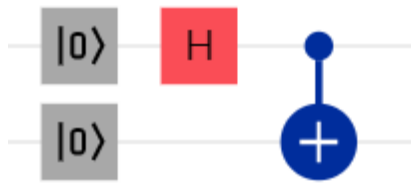

Figure 3.  After the CNOT gate on the right, an entangled state is formed: $\frac{1}{\sqrt{2}}(|0\rangle|0\rangle + |1\rangle|1\rangle)$.

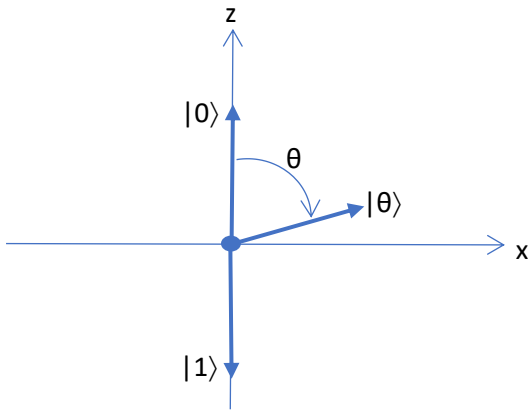

Figure 4. The xz-plane of the Bloch sphere. The qubit $|\theta\rangle = \cos(\theta/2)|0\rangle + \sin(\theta/2)|1\rangle$.

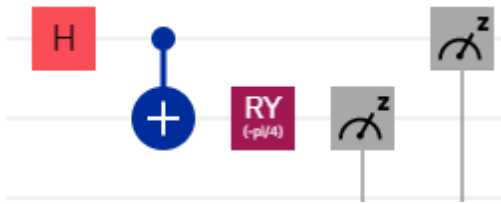

Figure 5. Circuit to determine $\langle A_1B_1\rangle$, with $\alpha_1 = 0$ and $\beta_1 = \pi/4$.

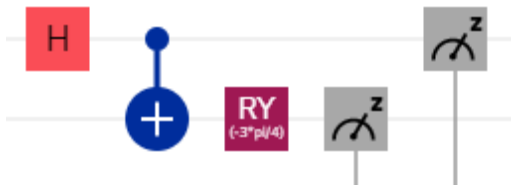

Figure 6. Circuit to determine $\langle A_1B_2\rangle$, with $\alpha_1 = 0$ and $\beta_2 = 3\pi/4$.

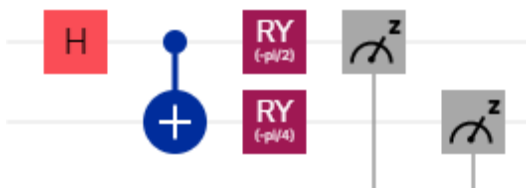

Figure 7. Circuit to determine $\langle A_2B_1\rangle$, with $\alpha_2 = \pi/2$ and $\beta_1 = \pi/4$.

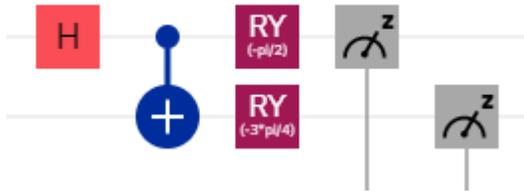

Figure 8. Circuit to determine <A$_2$B$_2$>, with $\alpha_2 = \pi/2$ and $\beta_2 = 3\pi/4$.